\documentclass[aps,prl,twocolumn,showpacs,superscriptaddress,groupedaddress]{revtex4}  
\usepackage{graphicx}  
\usepackage{dcolumn}   
\usepackage{bm}        
\usepackage{amssymb}   

\begin{document}
\title{Perpetual floating drops}

\author{Peter V. Pikhitsa}
\email{peter@snu.ac.kr}
\affiliation{Department of Mechanical and Aerospace Engineering,
Seoul National University, 08826 Seoul, Korea}

\begin{abstract}
The stability of floating drops on the liquid surface of  the  same
liquid is considered in terms of viscous drainage theory.  We  have
expressed the minimal thickness of the air film, separating the drop
from the liquid surface, and the  lifetime  of  the
drop  through  the Hamaker constant,  characterizing   the   intensity   of
van-der-Waals forces which make the air film unstable. It is  shown
that a horizontally moving or just rotating drop can have an infinite lifetime  when
the drop surface velocity exceeds some critical value. A spectacular example of a long-living rotating paraffin droplet on a burning candle is given.  Results obtained can
help to control a liquid surface {\it in situ} with  the  use  of
floating/rotating drops.
\end{abstract}


\date{\today}

\maketitle

\section{Introduction}
Floating drops have been studied for many years  (see  \cite{Baird}
and Refs. therein)  along  with  inverse  bubbles.  It  was  found
that {\it in vacuo} it was impossible to create a floating drop .
It confirms the responsibility of an air film, separating the
drop  from  the liquid surface,for the existence of the drop. This
air film is not that easy to remove because viscous air drainage
from between the drop and the liquid surface goes very slowly as
the film thickness $h$ decreases, enlarging considerably the
lifetime $\tau$ of the  drop up to tens of seconds.  Experiments
\cite{Baird}  have  proved  the existence  of   some   minimal   air
film   thickness  ($h_c\approx 3000 $ \AA ) for which the air film
becomes unstable.

The lifetime of small drops (diameter less than $0.5$cm) is restricted
mainly by $h_c$. These drops are  not  that  sensitive  to
random shocks because surface tension makes the drops rigid enough.  As
it was noticed in \cite{Baird} the value of $h_c$  is  determined  by
interatomic  quasimacroscopic  van-der-Waals  forces   \cite{Mahanty}
between the liquid boundaries of the air film, but $h_c$  was  not
obtained.

Here  we  show  that  van-der-Waals  forces,  pulling
the two liquid surfaces of the air film together and therefore  working
against surface tension, make the  liquid  surface unstable (when
strong enough) and this can happen  only  for  sufficiently  small
film thickness $h\leq h_c$. The mechanism  of  instability  is
similar to the dropping from a wide  ceiling  covered  with a water
layer where gravitational acceleration $g$ plays the same role as van-
der-Waals forces does. We express $h_c$ through the  van-der-Waals
constant (see below)  and  thus  link  macroscopically  measurable
lifetime  of  a  floating  drop  with  quasimacroscopic  parameter
describing  interaction  between  atoms  of   a liquid.   A   strong
stabilizing  effect  takes  place  for   floating   drops   moving
horizontally along the liquid surface. We obtain the critical
velocity $V_c$ above which the drops have  an  infinite  lifetime.
$V_c$ is found to be several cm/sec. We illustrate the drop stability with an experiment where a liquid paraffin drop lies on the surface of liquid paraffin indefinitely long, while rotating with surface velocity above $V_c$ under the temperature gradient of the candle flame. \\

\section{The viscous drainage theory of a floating drop}

For small drops we assume the following simplifications which help
us to grasp the principal features of the  phenomena  in a simplest
way:

$\left(i\right)$ the surface tension is strong enough to consider
the liquid surfaces of the air film rigid \cite{Baird};

$\left({ii}\right)$ the film thickness keeps approximately  the
same  along  the film;

$\left({iii}\right)$ the air film is considered flat which  slightly
alters  some coefficients but does not change results in principal.

The assumptions made are  valid  because  the  film  thickness  is
negligibly small in comparison with the radius $R$ of the drop.
As is known, the force of viscous friction between  two  rigid  flat
round plates being pulled  together  with  the  velocity
$v=\frac{dh}{dt}$  which prevents their soon  contact  reads \cite{Landau} :
 \begin{equation}
F={{3{\pi} v  {\eta}  R_f^4}  \over
{2h^3}}\mbox{,}
\end{equation}                                 
where $\eta$ is the air viscosity, $R_f$  is  the  radius  of  the
plate. In our case $R_f$ means the radius of the air film and $v$
means the drop velocity down to the underlying liquid surface. The
Newton equation of motion for the floating drop takes the form
\begin{equation}\frac{d^2h}{dt^2}=-g-{{\gamma}\over{h^3}}{{dh}
\over{dt}}\mbox{,}\end{equation}                             
where
\begin{equation}
{\gamma}={{3{\pi} {\eta}  R_f^4}  \over
{2m}}\mbox{,}
\end{equation}                                 
$m={\frac{4{\pi}}{3}}R^3{\rho}$ is the mass of the drop, ${\rho}$ is the liquid
density. We introduce  the  parameter  ${\beta}=\frac{R_f}{R}$  to
rewrite Eq.(3) as
\begin{equation}
{\gamma}={\frac{9}{8}}{{{\eta} {\beta}^4R}  \over
{\rho}}\mbox{.}
\end{equation}                                 

For a real floating drop ${\beta}$ may depend on $R$. As soon  as  the
drop lifetime is much larger than inverse  capillary  frequencies,
one can consider  the  drop  and  the  liquid  surface  to  be  in
equilibrium. Comparing  the  force  of  surface  tension  (Laplace
pressure of the curved surface) to the  weight  of  the  drop  one
obtains the equilibrium condition
\begin{equation}
{\frac{\sigma}{R}}{\int_0^{\beta}2{\pi}R
cos({\theta})sin({\theta})Rd{\theta}}=mg
\end{equation} 
which leads to
\begin{equation}
sin({\beta})= 2\sqrt{\frac{{\rho}g}{3{\sigma}}}R
\end{equation} 
so that for very small drops ${\beta}$  is  proportional  to  $R$.

One can see that Eq.(2) has an asymptotic solution (large time)
\begin{equation}
h=\sqrt{\frac{\gamma}{2gt}}=
{\frac{3}{4}}{\sqrt{{{\eta}R}\over{{\rho}gt}}}{\beta}
\end{equation} 
which slightly differs from the corresponding formula in  \cite{Baird}
by a dimensionless coefficient. According to Eq.(7) $h$ would slowly
go down to zero and this would take an infinite time had $h_c$ not
existed. \\

\section{The van-der-Waals instability of the air film}

The attraction van-der-Waals force per unit area between  two
liquid  surfaces with the separation distance $h$  in
non-retarded  region  (small distances) is known to be \cite{Mahanty}
\begin{equation}P(h)=-\frac{1}{6{\pi}}
\frac{A}{h^3}\mbox{,}\end{equation}                
where $A$ is the Hamaker constant which is simply  connected  with
the microscopic constant of atomic interaction $C_6$:
\begin{equation} A={\pi}^{2}n_{1}n_{2}C_{6}\mbox{,} \end{equation}   
where $n_1$,$n_2$ are the particle densities of  attracting  bodies
and $C_6$ determines the attraction energy $U=-{\frac{C_6}{R^6}}$  in
non-retarded region. Values of $A$ vary widely  from  ${10}^{-14}$ erg to
${10}^{-12}$ erg for different substances. For  large  $h$  the  power
index of $h$ alters so that
\begin{equation}P(h)=-\frac{B}{h^4}.\end{equation}    
The value of $B$ is of order of ${10}^{-19}$ erg cm \cite{Baird}.

The      force      $F(h)$      creates      an       acceleration
$g_f(h)=\frac{1}{\rho}\frac{dP(h)}{dh}$ which  shifts  the  frequency
${\omega}(k)$ of capillary waves with the wave number $k$ \cite{Landau}:
\begin{equation}{\omega}^2(k)=
\frac{\sigma}{\rho}k^3-g_f(h)k\end{equation}            
An instability  occurs  when  ${\omega}(k)=0$  which  defines  the
critical wave number
\begin{equation}
k_c=\sqrt{{{\rho}{g_f(h)}}\over{\sigma}}
\end{equation}
The instability  of  the  air  film  starts  developing  when  the
critical  wave  length  ${\lambda}_c=\frac{2{\pi}}{k_c}$   becomes
equal to the air film diameter:
\begin{equation} 2R_f={\lambda}_c\end{equation}

From Eq.(12) and Eq.(13) it is easy to obtain $h_c$,  choosing  an
appropriate form of $g_f(h)$, using either Eq.(9) or Eq.(10).  The
experimental \cite{Baird} showed $h_c$ to be in the retarded  region,  so
we choose here Eq.(10). Combining  Eqs.(10),(12),(13)  and  Eq.(7)
one gets
\begin{equation}
h_c={\left({4B}\over{{\pi}^2{\sigma}}\right)}^{\frac{1}{5}}
R^{\frac{2}{5}}{\beta}^{\frac{2}{5}}
\end{equation}

\begin{equation}
{\tau}_c={\frac{9}{16}}{\frac{\eta}{{\rho}g}}
{\left({\frac{\sigma}{4B}}\right)}^{\frac{2}{5}} R^{\frac{1}{5}}{\beta}^
{\frac{6}{5}}
\end{equation}
Substitution of parameters
${\eta}\approx{2}{\cdot}10^{-4} erg sec/{cm^{\small{3}}}, {\sigma}\approx 68
erg /cm^{\small{2}}$ (water) and $R\approx 0.2 $cm  with
${\beta  }\sim  1$  into   Eqs.(14),(15)   gives   rough
estimates
\begin{displaymath}h_c\sim 5{\cdot} 10^{-4}\mbox{cm}\end{displaymath}
\begin{displaymath}{\tau}_c\sim 10\mbox{sec}\end{displaymath}
which do not contradict the experimental.
It is worth mentioning that Eq.(15) supports the  observation  that
for the small drops of different sizes lifetimes are roughly
proportional  to  their  radii.  Eq.(15)   predicts   ${\tau}_c\sim
R^{\frac{7}{5}}$ (for ${\beta}\sim R$), and for  very  small  drops
one  should  choose  Eq.(9)  instead  of  Eq.(10)  which  leads  to
$\tau\sim R$ exactly. \\

\section{The critical velocity of horizontally moving drops}

\begin{figure}[h]
\includegraphics[width=0.5\textwidth]{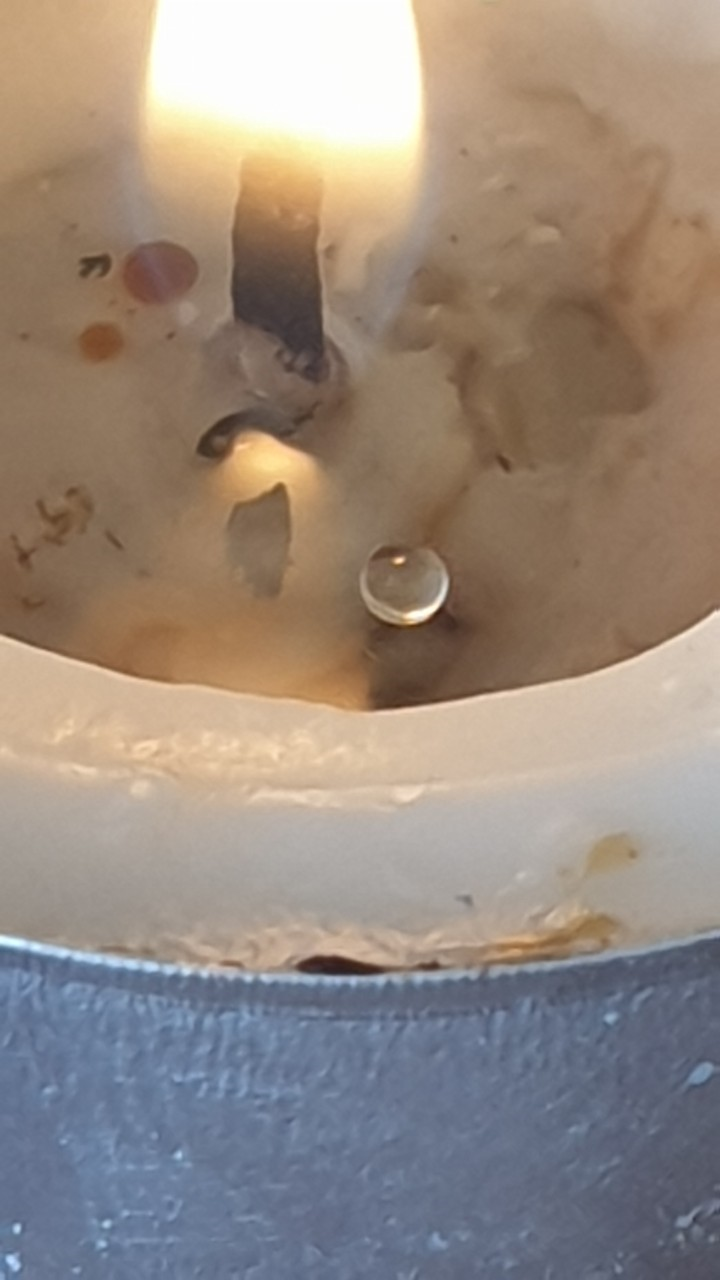}
\caption{{\em A long-living liquid paraffin droplet floating near a candle flame.} }
\label{fig:drp}
\end{figure}

As  was  mentioned  in  \cite{Baird}  the  presence  of  an  ionizing
surface-active  agents  does  not  greatly   affect   the   minimum
coalescence thickness and the drainage theory is valid,  taking  no
account of electrostatic  repulsion.  Nevertheless  there  exist  a
rather strong stabilizing effect within the viscous drainage theory
of floating drops. When a floating drop  moves  along  the  surface
there appears a viscous force which can prevent the film  thickness
from approaching the  critical  value  $h_c$  where  the  film  may
collapse.  We  rewrite  Eq.(2),   adding   the   viscous   pressure
${\eta}{\frac{V_d}{h}}$, into a form
\begin{equation}
\frac{d^2h}{dt^2}=-g-{{\gamma}\over{h^3}}
{{dh}\over{dt}}+ {\eta}{\frac{V_d}{h}}\frac{{\pi}R^{2}_{f}}{m}\mbox{,}
\end{equation} 
where $V_d$  is  the  horizontal  velocity  of  the  drop.  The  film
thickness of the stationary moving drop is
\begin{equation}h={{{\eta}V_d{\pi}R_f^2}\over{mg}}.\end{equation}
For $h\geq h_c$ the drop has an  infinite  lifetime.  The  velocity
threshold $V_c$ can be obtained from Eq.(17) and Eq.(14) and reads
\begin{equation}
V_c=\frac{4{\rho}g}{3{\eta}}
{\left(\frac{4B}{{\pi}^2{\sigma}}\right)}^{\frac{1}{5}} R^{\frac{7}{5}}  {\beta}^
{-\frac{8}{5}}.
\end{equation}
Numeric estimates lead to the value $V_c\sim 10 $cm/sec .
One can see that this value  is  quite  reasonable  for  frequently
observable drops dispersed on the sea surface. Our experiment  with
a liquid paraffine drop, floating  and  quickly  rotating  near  the
candle flame because of the nonuniform heating (see Fig.~\ref{fig:drp} and Supplemental Material at [URL will be inserted by publisher] for [a movie of the floating paraffin droplet on a candle]), also  confirms  the
numerical estimations. The rotation surface velocity (which exceeds the critical velocity $V_c$)can be found by measuring the drop angular velocity while visualizing the drop rotation with the speckles of soot which accidentally contaminate the drop.

\section{Summary}

Using the viscous drainage  theory  and  taking  into  account  the
van-der-Waals forces we link together macroscopic parameters of the
floating drop ${\tau}_c$ and $V_c$ with quasimicroscopic  parameter
$B$, which characterizes van-der-Waals forces. This means  that  the
floating drop may be used as a tool to  control  the  intensity  of
molecular interactions between the liquid  surfaces  {\it  in  situ},
which could be important  to control the presence of
surface-active agents.\\

\end{document}